\documentclass[twocolumn,twoside,slac_two]{revtex4}
\usepackage{graphicx}
\usepackage{fancyhdr}
\def\Dzkpi{D^0\to K^-\pi^+}
\def\Dzkpipiz{D^0\to K^-\pi^+\pi^0}
\def\Dzkpipipi{D^0\to K^-\pi^+\pi^+\pi^-}
\def\Dpkpipi{D^+\to K^-\pi^+\pi^+}
\def\Dpkpipipiz{D^+\to K^-\pi^+\pi^+\pi^0}
\def\Dpkspi{D^+\to K^0_S\pi^+}
\def\Dpkspipiz{D^+\to K^0_S\pi^+\pi^0}
\def\Dpkspipipi{D^+\to K^0_S\pi^+\pi^+\pi^-}
\def\Dpkkpi{D^+\to K^+K^-\pi^+}

\newcommand{\Dp}{D^+}
\newcommand{\Dm}{D^-}
\newcommand{\Dz}{D^0}
\newcommand{\Dbar}{\overline{D}}
\newcommand{\Dzbar}{\overline{D}{}^0}
\newcommand{\calB}{\mathcal{B}}

\newcommand{\Km}{K^-}

\newcommand{\pip}{\pi^+}

\begin{document}

\title{Determination of Charm Hadronic Branching Fractions at CLEO-c}

%

\author{A. Ryd Representing the CLEO Collaboration}
\affiliation{Laboratory for Elementary-Particle Physics, Cornell University, Ithaca NY 14853, USA}

\begin{abstract}
Recent results from CLEO-c on measurements of absolute 
hadronic branching 
fractions of $D^0$, $D^+$, and $D_s^+$ mesons are presented. 
\end{abstract}

\maketitle

\thispagestyle{fancy}

\section{Introduction}

 Precise measurements of absolute hadronic branching fractions
for $D^0$, $D^+$, and $D_s^+$ meson decays are important as they serve
to normalize most $B$ and $B_s$ decays as well as many
charm decays.  

Results from the CLEO-c experiment at the Cornell 
Electron Positron Storage Ring
based on 281 pb$^{-1}$ recorded at the $\psi(3770)$ are
presented here for studies of $D^0$ and $D^+$ decays. 
In addition, CLEO-c has analyzed 298 pb$^{-1}$
of $e^+e^-$ annihilation data near $E_{\rm cm}=4170$ MeV for
studies of $D_s$ decays.
These samples
provide very clean environments for studying decays of 
$D$ and $D_s$ mesons. 
The $\psi(3770)$,
produced in the $e^+e^-$ annihilation, decays to pairs of $D$ mesons,
either $D^+D^-$ or $D^0\bar D^0$. In particular, the
produced $D$ mesons can not be accompanied
by any additional pions. At $E_{\rm cm}=4170$ MeV $D_s$ 
mesons are primarily produced as $D_s^{+}D_s^{*-}$ and
$D_s^{*+}D_s^{-}$ pairs. 

First, I will discuss the determination of the absolute hadronic
$D^0$, $D^+$, and $D_s^+$ branching fractions. Then I will
present CLEO-c measurements of inclusive $\eta$, $\eta'$,
and $\phi$ decays; the doubly Cabibbo suppressed decay
$D^+\to K^+\pi^0$; studies of $D\to K_S\pi$ and $D\to K_L\pi$;
$D_s$ decays to two pseudoscalars; and two-body $D^0$ and
$D^+$ decays to pairs of kaons.

\section{Absolute $D^0$ and $D^+$ hadronic branching fractions}

This analysis~\cite{dhadprd} makes use of a 'double tag' technique
initially used by Mark III~\cite{markiii}.
In this technique the yields of single tags, where
one $D$ meson is reconstructed, and
double tags, where both $D$ mesons are reconstructed,
are determined.
The number of reconstructed single tags,
separately for $D$ and $\bar D$ decays, are given by
$N_i=\epsilon_i{\cal B}_i N_{D\bar D}$ and 
${\bar N}_j=\bar \epsilon_j{\cal B}_j N_{D\bar D}$, respectively,
where $\epsilon_i$ and ${\cal B}_i$ are the efficiency
and branching fraction for mode $i$. Similarly,
the number of double tags reconstructed are given
by $N_{ij}=\epsilon_{ij}{\cal B}_i{\cal B}_j N_{D\bar D}$
where $i$ and $j$ label the $D$ and $\bar D$ mode used
to reconstruct the event and $\epsilon_{ij}$ is the
efficiency for reconstructing the final state.
Combining the equations above and solving for $N_{D\bar D}$
gives the number of produced $D\bar D$ events 
as 
$$
N_{D\bar D}={{N_i}{\bar N_j}\over N_{ij}}{\epsilon_{ij}\over \epsilon_i\bar\epsilon_j}
$$
and the branching fractions 
$$
{\cal B}_i={N_{ij}\over N_j}{{\epsilon_j}\over \epsilon_{ij}}.
$$
In this analysis we determine all the
single tag and double tag yields in data, determine the efficiencies
from Monte Carlo simulations of the detector
response, and extract the branching
fractions and $D\bar D$ yields from a combined fit~\cite{brfrfitter} to all 
measured data yields.

\begin{figure*}[tb]
\begin{center}
\includegraphics[width=0.75\linewidth]{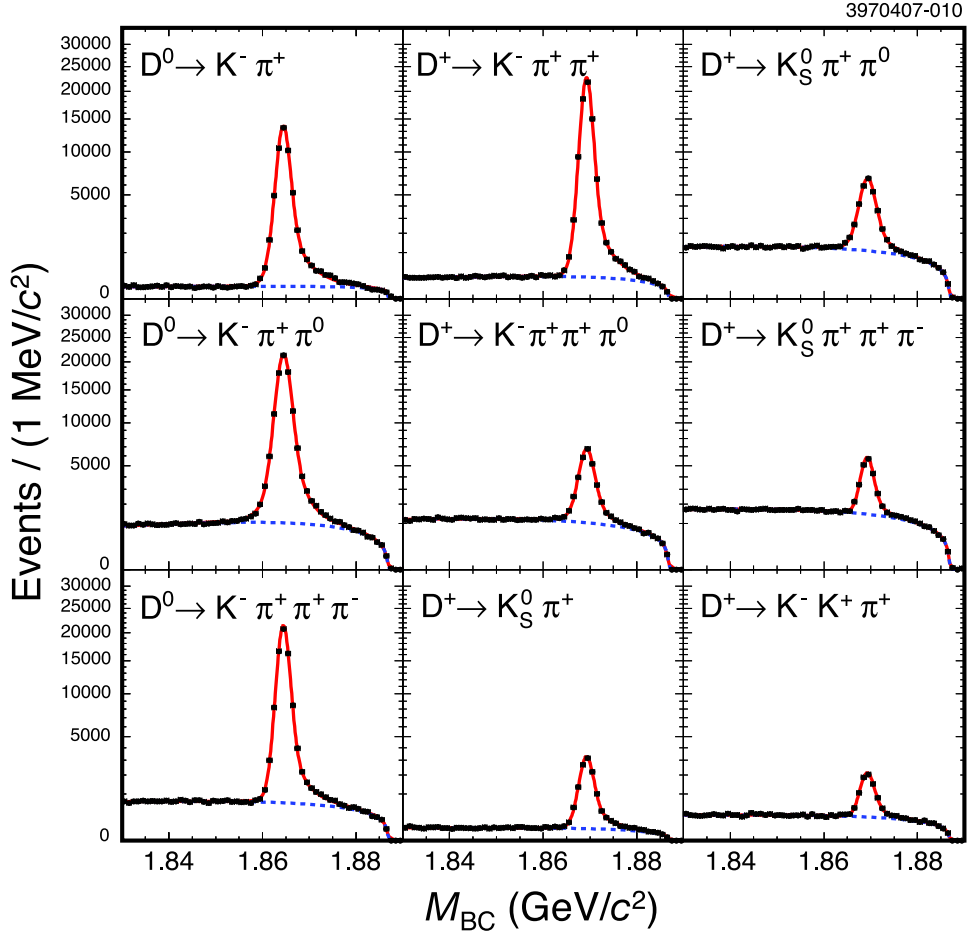}
\caption{The fits for the single tag yields. The background is 
described by the ARGUS threshold function and the signal shape
includes the effects of beam energy spread, momentum resolution,
initial state radiation, and the $\psi(3770)$ lineshape. }
\label{fig:dhad_st}
\end{center}
\end{figure*}

\begin{figure*}[tb]
\begin{center}
\includegraphics[width=0.49\linewidth]{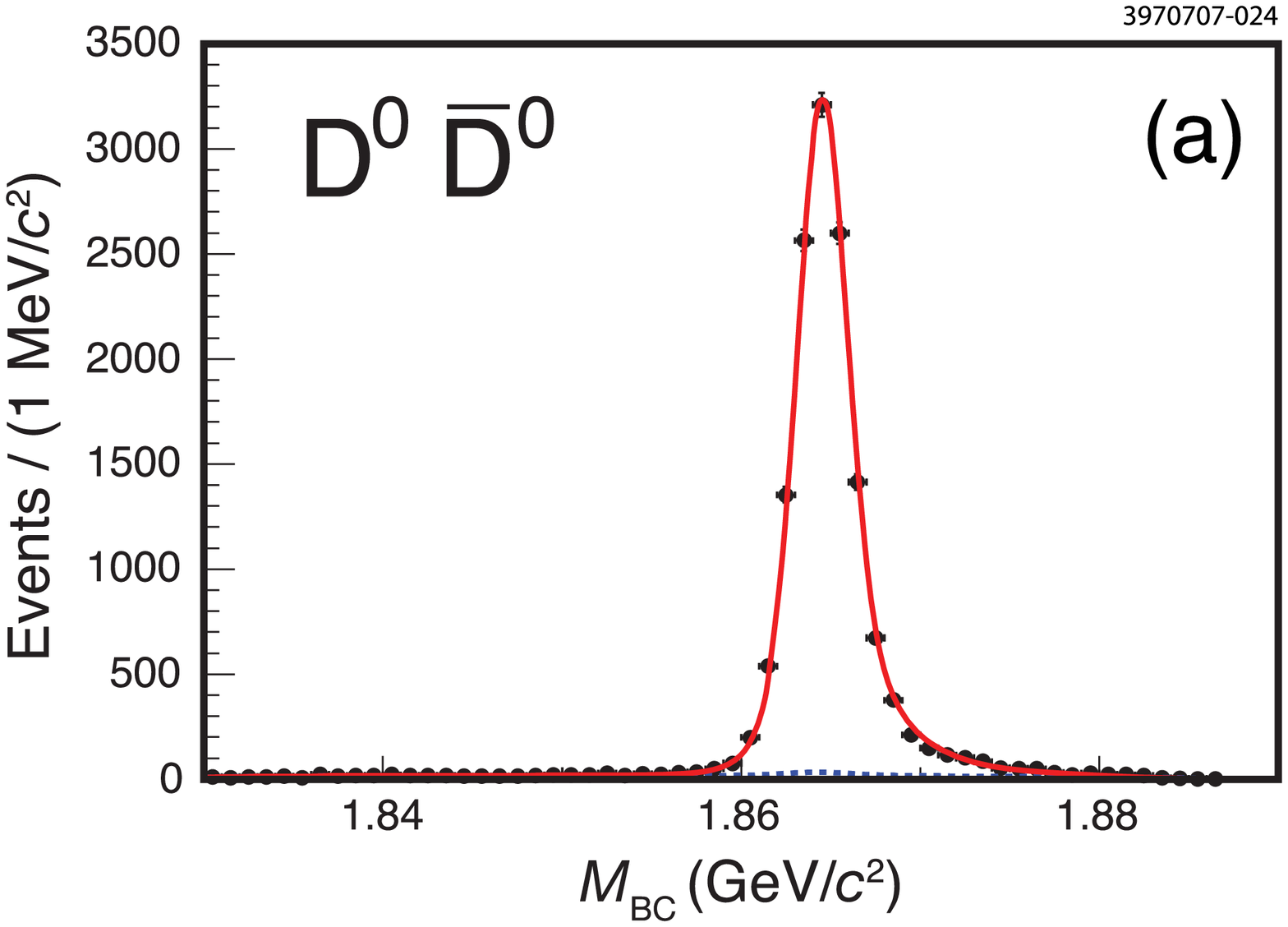}
\includegraphics[width=0.49\linewidth]{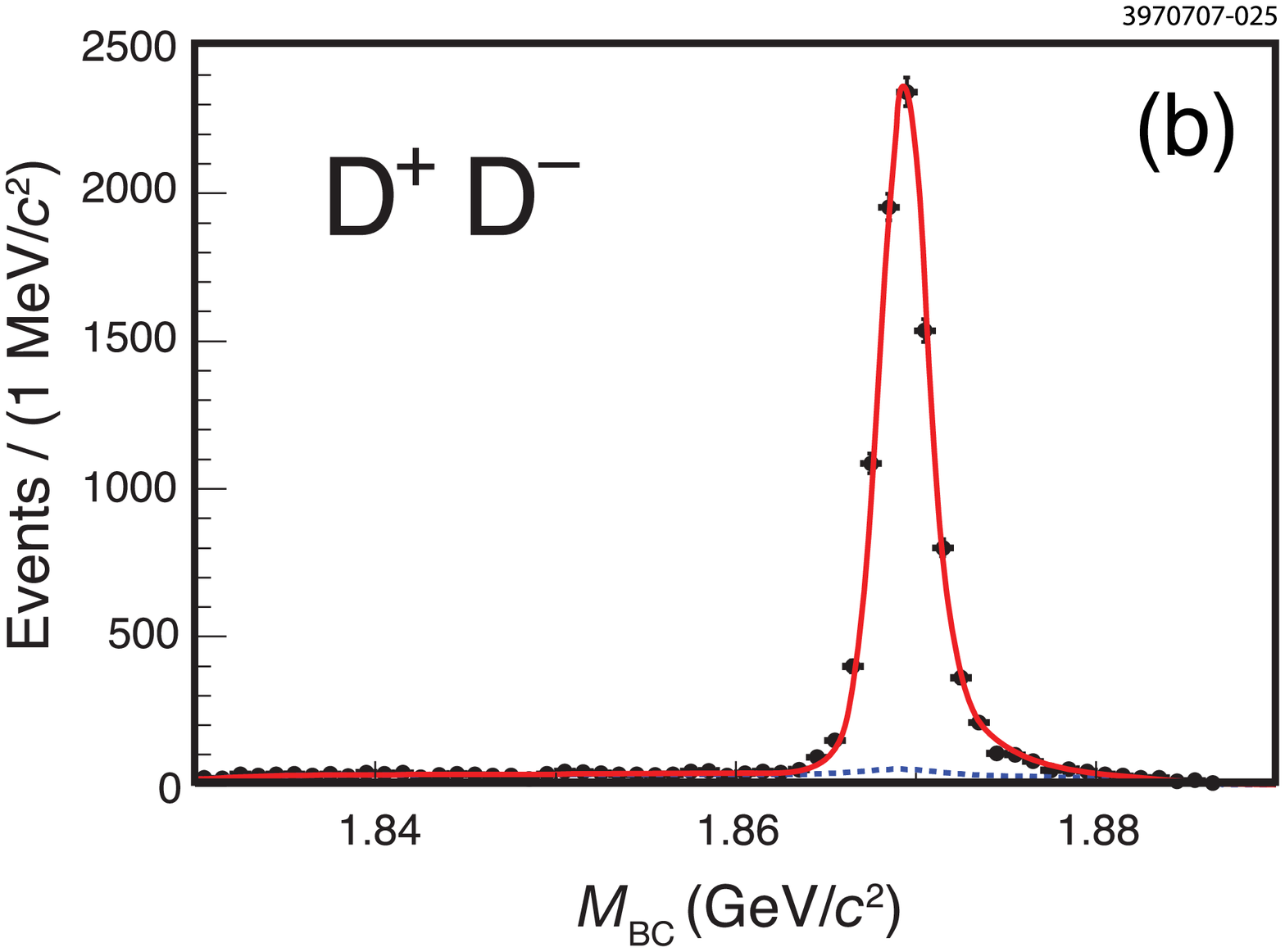}
\caption{The fit for the double tag yields combined over
all modes for charged and neutral modes separately.}
\label{fig:dhad_dt}
\end{center}
\end{figure*}

This analysis uses three $D^0$ decay modes 
($D^0\to K^-\pi^+$, $D^0\to K^-\pi^+\pi^0$, and $D^0\to K^-\pi^+\pi^-\pi^+$) 
and six $D^+$ decay modes
($D^+\to K^-\pi^+\pi^+$, $D^+\to K^-\pi^+\pi^+\pi^0$, $D^+\to K^0_S\pi^+$, $D^+\to K^0_S\pi^+\pi^0$,
 $D^+\to K^0_S\pi^+\pi^-\pi^+$, and $D^+\to K^-K^+\pi^+$).
The single tag yields are shown in Fig.~\ref{fig:dhad_st}.
The combined double tag
yields are shown in Fig.~\ref{fig:dhad_dt}
for charged and neutral $D$ modes separately.
The scale of the statistical errors on the branching fractions
are set by the number of double tags and  
precisions of 
$\approx 0.8\%$ and $\approx 1.0\%$ are obtained for the neutral and charged modes
respectively.
The branching fractions obtained are summarized in Table~\ref{tab:dhadresults}
\footnote{The result presented here represents the final results and are 
slightly different from the results presented at the workshop.}.
For the branching fractions we quote three uncertainties. The first is the statistical
uncertainty, the second is the systematic uncertainties excluding the uncertainty
in the modeling of final state radiation (FSR), and the third error is the FSR
uncertainty. For the $D^0\to K^-\pi^+$ mode the effect of the FSR is a 3.0\% correction.
We have taken the uncertainty of the FSR correction to be about 30\% of the
correction. This covers the difference between including or excluding the effect
of interference in simulating FSR in the decay $D^0\to K^-\pi^+$.

\begin{table*}[htb]
\caption{Fitted branching fractions and $D\Dbar$ pair
yields.  For $N_{\Dz\Dzbar}$ and $N_{\Dp\Dm}$, uncertainties are statistical and systematic,
respectively.  For branching fractions and ratios, the systematic uncertainties are 
divided into the contribution from FSR (third uncertainty) and all others 
combined (second uncertainty).  The column of fractional systematic errors combines 
all systematic errors, including FSR.  The last column, $\Delta_{\rm FSR}$, is the 
relative shift in the fit results when FSR is not included in the Monte Carlo simulations 
used to determine efficiencies.\label{tab:dhadresults}}
\begin{tabular}{lcccc} \hline\hline
Parameter & Fitted Value & \multicolumn{2}{c}{Fractional Error} & $\Delta_{\rm FSR}$\\[-0.6ex]
&& Stat.(\%) & Syst.(\%)  &  (\%) \\ \hline

$N_{\Dz\Dzbar}$	 & $(1.031 \pm 0.008 \pm 0.013)\times 10^6$	 & $0.8$	 & $1.3$	 & $+0.1$\\
${\cal B}(\Dzkpi)$	 & $(3.891 \pm 0.035 \pm 0.059 \pm 0.035)\%$	 & $0.9$	 & $1.8$	 & $-3.0$\\
${\cal B}(\Dzkpipiz)$	 & $(14.57 \pm 0.12 \pm 0.38 \pm 0.05)\%$	 & $0.8$	 & $2.7$	 & $-1.1$\\
${\cal B}(\Dzkpipipi)$	 & $(8.30 \pm 0.07 \pm 0.19 \pm 0.07)\%$	 & $0.9$	 & $2.4$	 & $-2.4$\\
$N_{\Dp\Dm}$	 & $(0.819 \pm 0.008 \pm 0.010)\times 10^6$	 & $1.0$	 & $1.2$	 & $+0.1$\\
${\cal B}(\Dpkpipi)$	 & $(9.14 \pm 0.10 \pm 0.16 \pm 0.07)\%$	 & $1.1$	 & $1.9$	 & $-2.3$\\
${\cal B}(\Dpkpipipiz)$	 & $(5.98 \pm 0.08 \pm 0.16 \pm 0.02)\%$	 & $1.3$	 & $2.8$	 & $-1.0$\\
${\cal B}(\Dpkspi)$ 	 & $(1.526 \pm 0.022 \pm 0.037 \pm 0.009)\%$	 & $1.4$	 & $2.5$	 & $-1.8$\\
${\cal B}(\Dpkspipiz)$	 & $(6.99 \pm 0.09 \pm 0.25 \pm 0.01)\%$	 & $1.3$	 & $3.5$	 & $-0.4$\\
${\cal B}(\Dpkspipipi)$	 & $(3.122 \pm 0.046 \pm 0.094 \pm 0.019)\%$	 & $1.5$	 & $3.0$	 & $-1.9$\\
${\cal B}(\Dpkkpi)$	 & $(0.935 \pm 0.017 \pm 0.024 \pm 0.003)\%$ & $1.8$	 & $2.6$	 & $-1.2$\\ \hline
${{\calB}(\Dzkpipiz)}/{{\calB}(\Km\pip)}$	 & $3.744 \pm 0.022 \pm 0.093 \pm 0.021$	 & $0.6$	 & $2.6$	 & $+1.9$\\
${{\calB}(\Dzkpipipi)}/{{\calB}(\Km\pip)}$	 & $2.133 \pm 0.013 \pm 0.037 \pm 0.002$	 & $0.6$	 & $1.7$	 & $+0.5$\\
${{\calB}(\Dpkpipipiz)}/{{\calB}(\Km\pip\pip)}$	 & $0.654 \pm 0.006 \pm 0.018 \pm 0.003$	 & $0.9$	 & $2.7$	 & $+1.4$\\
${{\calB}(\Dpkspi)}/{{\calB}(\Km\pip\pip)}$	 & $0.1668 \pm 0.0018 \pm 0.0038 \pm 0.0003$	 & $1.1$	 & $2.3$	 & $+0.5$\\
${{\calB}(\Dpkspipiz)}/{{\calB}(\Km\pip\pip)}$	 & $0.764 \pm 0.007 \pm 0.027 \pm 0.005$	 & $0.9$	 & $3.5$	 & $+2.0$\\
${{\calB}(\Dpkspipipi)}/{{\calB}(\Km\pip\pip)}$	 & $0.3414 \pm 0.0039 \pm 0.0093 \pm 0.0004$	 & $1.1$	 & $2.7$	 & $+0.4$\\
${{\calB}(\Dpkkpi)}/{{\calB}(\Km\pip\pip)}$	 & $0.1022 \pm 0.0015 \pm 0.0022 \pm 0.0004$	 & $1.5$	 & $2.2$	 & $+1.1$\\
\hline\hline
\end{tabular}
\end{table*}

\section{Absolute branching fractions for hadronic $D_s$ decays}

This analysis uses a sample of 298 pb$^{-1}$ of data recorded at
a center-of-mas energy of 4170 MeV. At this energy $D_s$ mesons
are produced, predominantly, as $D_s^+D_s^{*-}$ or $D_s^-D_s^{*+}$
pairs. We use the same tagging technique as for the hadronic
$D$ branching fractions; we reconstruct samples of single tags and
double tags and use this to extract the branching fractions. 

\begin{figure*}[tb]
\begin{center}
\includegraphics[width=0.75\linewidth]{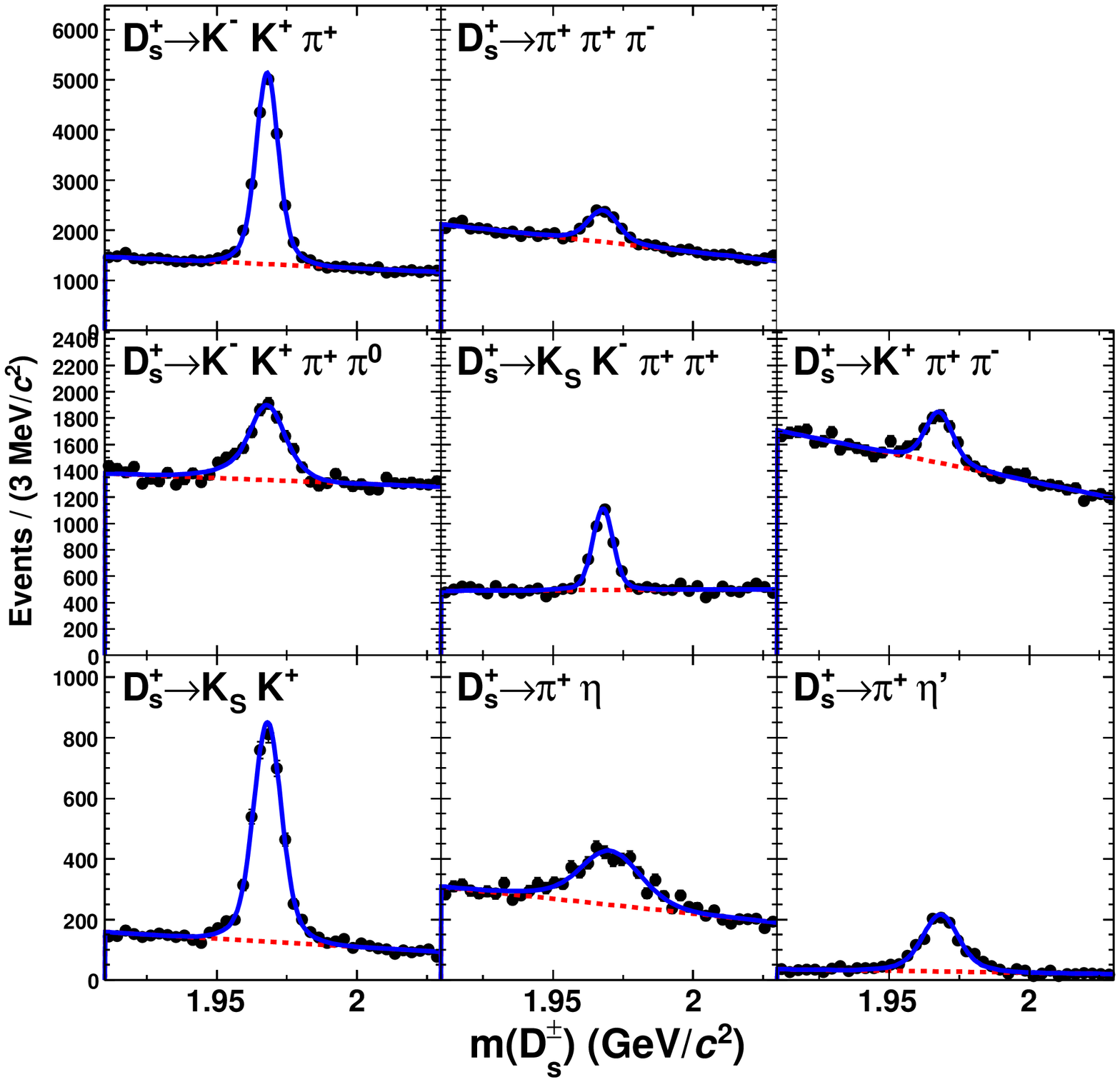}
\caption{Single tag yields for the reconstucted $D_s$ modes used in the 
analysis of the absolute hadronic $D_s$ branching fractions.}
\label{fig:cleoc_ds_st}
\end{center}
\end{figure*}

\begin{figure}[tb]
\begin{center}
\includegraphics[width=0.95\linewidth]{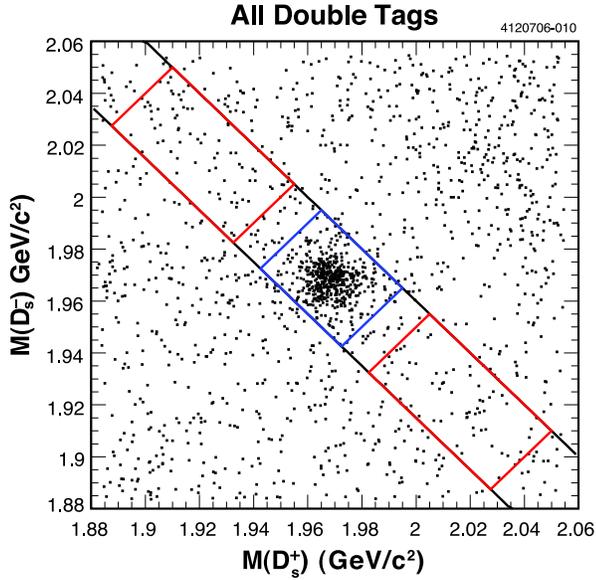}
\caption{The distribution of the reconstructed invariant mass of the $D_s^-$
candidate versus the $D_s^+$ candidate for the double tag candidates in
the absolute $D_s$ hadronic branching fraction analysis.}
\label{fig:cleoc_ds_dt}
\end{center}
\end{figure}

In this study eight $D_s$ final states are used
($D^+_s\to K^0_S K^+$,
$D^+_s\to K^+K^-\pi^+$,
$D^+_s\to K^+K^-\pi^+\pi^0$,
$D^+_s\to K^0_SK^-\pi^+\pi^+$,
$D^+_s\to \pi^+\pi^-\pi^+$,
$D^+_s\to \eta\pi^+$,
$D^+_s\to \eta'\pi^+$, and
$D^+_s\to K^+\pi^-\pi^+$
). The single tag event yields are shown
in Fig.~\ref{fig:cleoc_ds_st}. The double tag yields are extracted
by a cut-and-count procedure in the plot of the invariant mass
of the $D^+_s$ vs. $D^-_s$. This plot is shown in Fig.~\ref{fig:cleoc_ds_dt}.
Backgrounds are subtracted from the sidebands indicated in
the plot and a total of $976\pm33$ double tag events are found. 

\begin{table*}[bt]
\small{
\caption{Preliminary branching fractions for $D_s$ decays determined in the 
CLEO-c analysis.
 }
\label{tab:cleoc_ds_brfr}
\begin{center}
\begin{tabular}{lccc}
\hline\hline
Mode & Branching Fraction (\%) & PDG 2006 fit (\%) \\ \hline
${\cal B}(D^+_s\to K^0_S K^+)$          & $1.56\pm0.08\pm0.05$ & $2.2\pm0.45$       \\
${\cal B}(D^+_s\to K^+K^-\pi^+)$        & $5.67\pm0.24\pm0.18$ & $5.2\pm 0.9$        \\
${\cal B}(D^+_s\to K^+K^-\pi^+\pi^0)$   & $5.58\pm0.29\pm0.45$ &         \\
${\cal B}(D^+_s\to K^0_SK^-\pi^+\pi^+)$ & $1.73\pm0.10\pm0.07$ & $2.65\pm0.7$        \\
${\cal B}(D^+_s\to \pi^+\pi^-\pi^+)$    & $1.13\pm0.07\pm0.05$ & $1.22\pm0.23$        \\
${\cal B}(D^+_s\to \eta\pi^+)$          & $1.63\pm0.11\pm0.17$ & $2.11\pm 0.35$        \\
${\cal B}(D^+_s\to \eta'\pi^+)$         & $3.98\pm0.26\pm0.32$ & $4.5\pm 0.7$      \\
${\cal B}(D^+_s\to K^+\pi^+\pi^-)$      & $0.71\pm0.05\pm0.03$ & $0.66\pm0.14$    \\ 
\hline\hline
\end{tabular}
\end{center}
}
\end{table*}

From these yields we determine the preliminary branching fractions
listed in Table~\ref{tab:cleoc_ds_brfr}.
We do not quote branching fractions for $D_s^+\to \phi\pi^+$
as the $\phi$ signal is not well defined. In particular, 
the $\phi$ resonance
interferes with the $f_0$ resonance. Instead we report 
preliminary results for partial
branching fractions for $D_s^+\to K^+K^-\pi^+$ in restricted
invariant mass ranges of $m_{KK}$ near the $\phi$ resonance.
These partial branching fractions are summarized in 
Table~\ref{tab:cleoc_dstophipi}.

\begin{table*}[bt]
\small{
\caption{Preliminary partial branching fractions for $D_s^+\to K^+K^-\pi^+$ in
limited $m(K^-K^+)$ ranges around the $\phi(1020)$ mass.
 }
\label{tab:cleoc_dstophipi}
\begin{center}
    \begin{tabular}{cc} \hline\hline
    $m(K^-K^+)$ range             &  Partial branching fraction($\%$) \\ \hline
    $|m(K^-K^+)-m_{\phi}|<\phantom{0}5$ MeV & $1.75 \pm 0.08 \pm 0.06 $  \\ 
    $|m(K^-K^+)-m_{\phi}|<10$ MeV & $2.07 \pm 0.10 \pm 0.05 $  \\ 
    $|m(K^-K^+)-m_{\phi}|<15$ MeV & $2.22 \pm 0.11 \pm 0.06 $  \\ 
    $|m(K^-K^+)-m_{\phi}|<20$ MeV & $2.32 \pm 0.11 \pm 0.06 $  \\  \hline\hline
    \end{tabular}
\end{center}
}
\end{table*}

\section{Inclusive measurements of $\eta$, $\eta'$, and $\phi$ production in $D$ and $D_s$ decays}

Using samples of tagged $D$ and $D_s$ decays CLEO-c has measured
the inclusive production of  $\eta$, $\eta'$, and $\phi$ mesons
by looking at the recoil against the tag~\cite{cleoc_inclusive}. The results are
summarized in Table~\ref{tab:cleoc_inclusive}. The knowledge of
inclusive measurements before this CLEO-c measurement was
poor, besides limits, only ${\cal B}(D^0\to \phi X)=(1.7\pm0.8)\%$ 
was measured. As expected the $\eta$, $\eta'$, and $\phi$ rates
are much higher in $D_s$ decays.

\begin{table}[bt]
\caption{Inclusive branching fractions of $D^0$, $D^+$ and $D_s^+$
meson decays to $\eta$, $\eta'$, and $\phi$.}
\label{tab:cleoc_inclusive}
\begin{center}
\begin{tabular}{lc}
\hline\hline
Decay  & ${\cal B}$ (\%) \\ \hline
$D^0\to\eta X$          & $9.5\pm0.4\pm0.8$        \\
$D^-\to\eta X$          & $6.3\pm0.5\pm0.5$         \\
$D_s^+\to\eta X$        & $23.5\pm3.1\pm2.0$         \\
\hline
$D^0\to\eta' X$         & $2.48\pm0.17\pm0.21$        \\
$D^-\to\eta' X$         & $1.04\pm0.16\pm0.09$         \\
$D_s^+\to\eta' X$       & $8.7\pm1.9\pm1.1$         \\
\hline
$D^0\to\phi X$          & $1.05\pm0.08\pm0.07$        \\
$D^-\to\phi X$          & $1.03\pm0.10\pm0.07$         \\
$D_s^+\to\phi X$        & $16.1\pm1.2\pm1.1$         \\
\hline\hline
\end{tabular}
\end{center}
\end{table}

\section{The doubly Cabibbo suppressed decay $D^+\to K^+\pi^0$}

CLEO-c~\cite{cleoc_dcsd} has reconstructed $D^+\to K^+\pi^0$ candidates in the
281 pb$^{-1}$ sample of $e^+e^-$ data recorded at the $\psi(3770)$. 
We find the branching fraction
${\cal B}(D^+\to K^+\pi^0)=(2.24\pm0.36\pm0.15\pm0.08)\times 10^{-4}$,
which is in good agreement with the recent BABAR measurement~\cite{babar_dcsd}  
${\cal B}(D^+\to K^+\pi^0)=(2.52\pm0.46\pm0.24\pm0.08)\times 10^{-4}$. 

\section{Modes with $K^0_L$ or $K^0_S$ in the final states}

It has commonly been assumed that $\Gamma(D\to K^0_S X)=\Gamma(D\to K^0_L X)$.
However, as pointed out by Bigi and Yamamoto~\cite{bigi} this is
not generally true as for many $D$ decays there are contributions from
Cabibbo favored and Cabibbo suppressed decays that interfere and
contributes differently to final states with $K^0_S$ and $K^0_L$.
As an example consider $D^0\to K^0_{S,L}\pi^0$. Contributions to
these final states involve the Cabibbo favored decay $D^0\to \bar K^0\pi^0$
as well as the Cabibbo suppressed decay $D^0\to K^0\pi^0$. However,
we don't observe the $K^0$ and the $\bar K^0$ but rather the $K^0_S$
and the $K^0_L$. As these two amplitudes interfere constructively to
form the $K^0_S$ final state we will see a rate asymmetry. Based
on factorization Bigi and Yamamoto predicted
\begin{eqnarray*}
R(D^0)&\equiv&{{\Gamma(D^0\to K^0_S\pi^0)-\Gamma(D^0\to K^0_L\pi^0)}\over
{\Gamma(D^0\to K^0_S\pi^0)+\Gamma(D^0\to K^0_L\pi^0)}}\\
     &\approx& 2\tan^2\theta_C\approx 0.11.\\
\end{eqnarray*}
Using tagged $D$ mesons CLEO-c has measured~\cite{cleoc_klks} this 
asymmetry and obtained
$$
R(D^0)=0.108\pm0.025\pm0.024,
$$
which is in good agreement with the prediction.

Similarly, CLEO-c has also measured the corresponding asymmetry in charged
$D$ mesons and obtained
\begin{eqnarray*}
R(D^+)&\equiv&{{\Gamma(D^+\to K^0_S\pi^+)-\Gamma(D^+\to K^0_L\pi^+)}\over
{\Gamma(D^+\to K^0_S\pi^+)+\Gamma(D^+\to K^0_L\pi^+)}}\\
   &=&0.022\pm0.016\pm0.018.\\
\end{eqnarray*}
Prediction of the asymmetry in charged $D$ decays is more 
involved. D.-N.~Gao~\cite{gao} predicts this asymmetry to be in the 
range 0.035 to 0.044, which is consistent with the observed asymmetry.

\section{$D_s$ decays to two pseudoscalars}

\begin{figure}[tb]
\begin{center}
\includegraphics[width=0.99\linewidth]{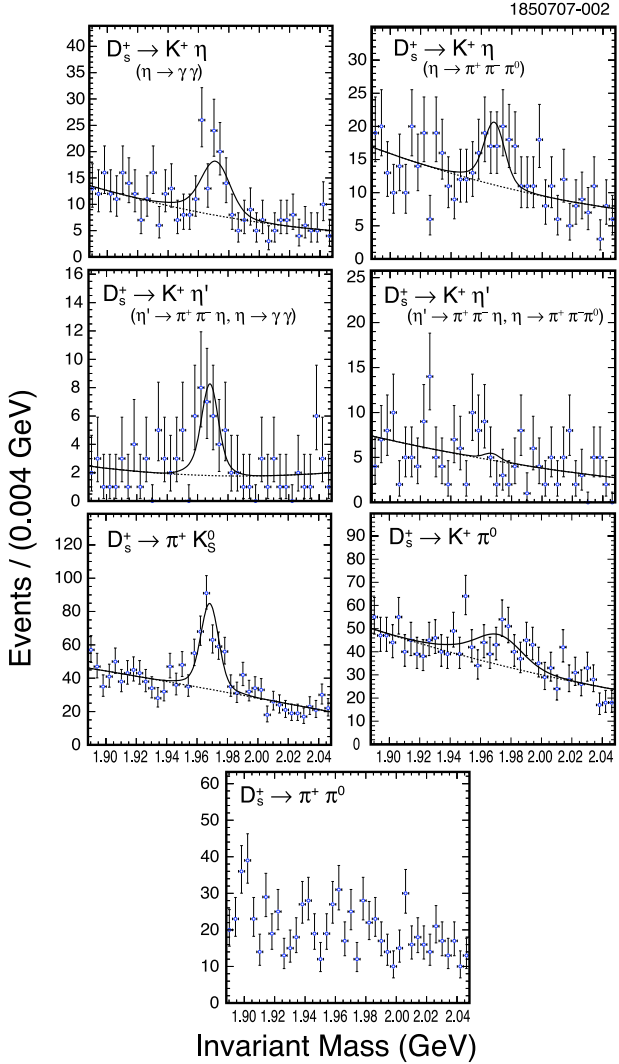}
\caption{Observed signals in the $D_s\to PP$ analysis.}
\label{fig:dstoPP}
\end{center}
\end{figure}

CLEO-c has performed a study of $D_s$ decays to a pair
of pseudoscalars. These final states consists of either a
$K^+$ or a $\pi^+$ and one of $\eta$, $\eta'$, $\pi^0$, or $K^0_S$.
In the analysis presented here the following final states are
studied: $D_s^+\to K^+\eta$, $D_s^+\to K^+\eta'$, $D_s^+\to K^+\pi^0$
$D_s^+\to \pi^+K^0_S$, and $D_s^+\to \pi^+\pi^0$. The final state
$D_s^+\to \pi^+\pi^0$ violates isospin and is expected to be
small. The details of the analysis can be found in Ref.~\cite{dstoPP}.
The signals are observed in the $D_s$ invariant mass distribution as peaks
at the $D_s$ mass. Significant signals are observed in all modes
except $D_s^+\to \pi^+\pi^0$. The observed mass distributions are
shown in Fig.~\ref{fig:dstoPP}. We measure the ratio of the
branching fractions of the Cabibbo suppressed modes with respect to 
the Cabibbo favored modes. The results are summarized in 
Table~\ref{tab:dstoPP}. The observed ratios of branching fractions
are consistent with the naive expectation of $|V_{cd}/V_{cs}|^2\approx 0.05$.
In addition, we have looked for a $CP$ asymmetry in rate for $D_s^+$
and $D_s^-$ decays. No evidence for any $CP$ asymmetry was found; the
results are summarized in Table~\ref{tab:dstoPPasym}.

\begin{table}[bt]
\caption{Branching ratios for the $D_s\to PP$ analysis.}
\label{tab:dstoPP}
\begin{center}
\begin{tabular}{lc}
\hline\hline
Mode  & ${\cal B}_S/{\cal B}_F$ (\%) \\ \hline
${\cal B}(D_s^+\to K^+\eta)/{\cal B}(D_s^+\to \pi^+\eta)$  & $8.9\pm1.5\pm0.4$ \\
${\cal B}(D_s^+\to K^+\eta')/{\cal B}(D_s^+\to \pi^+\eta')$  & $4.2\pm1.3\pm0.3$ \\
${\cal B}(D_s^+\to \pi^+K^0_S)/{\cal B}(D_s^+\to K^+K^0_S)$  & $8.2\pm0.9\pm0.2$ \\
${\cal B}(D_s^+\to K^+\pi^0)/{\cal B}(D_s^+\to K^+K^0_S)$  & $5.0\pm1.2\pm0.6$ \\
${\cal B}(D_s^+\to \pi^+\pi^0)/{\cal B}(D_s^+\to K^+K^0_S)$  & $<4.1\ (90\%\ {\rm CL})$\\ 
\hline\hline
\end{tabular}
\end{center}
\end{table}

\begin{table}[bt]
\caption{$CP$ asymmetries for Cabibbo suppressed $D_s\to PP$ decays.}
\label{tab:dstoPPasym}
\begin{center}
\begin{tabular}{lc}
\hline\hline
Mode  & $({\cal B}_+-{\cal B}_-)/({\cal B}_++{\cal B}_-)$ (\%) \\ \hline
${\cal A}(D_s^+\to K^+\eta)$ & $-20\pm18$ \\
${\cal A}(D_s^+\to K^+\eta')$ & $-17\pm37$ \\
${\cal A}(D_s^+\to \pi^+K^0_S)$ & $27\pm11$ \\
${\cal A}(D_s^+\to K^+\pi^0)$ & $2\pm29$  \\
\hline\hline
\end{tabular}
\end{center}
\end{table}

\section{$D^0$ and $D^+$ decays to two kaons}

\begin{figure*}[tb]
\begin{center}
\includegraphics[width=0.32\linewidth]{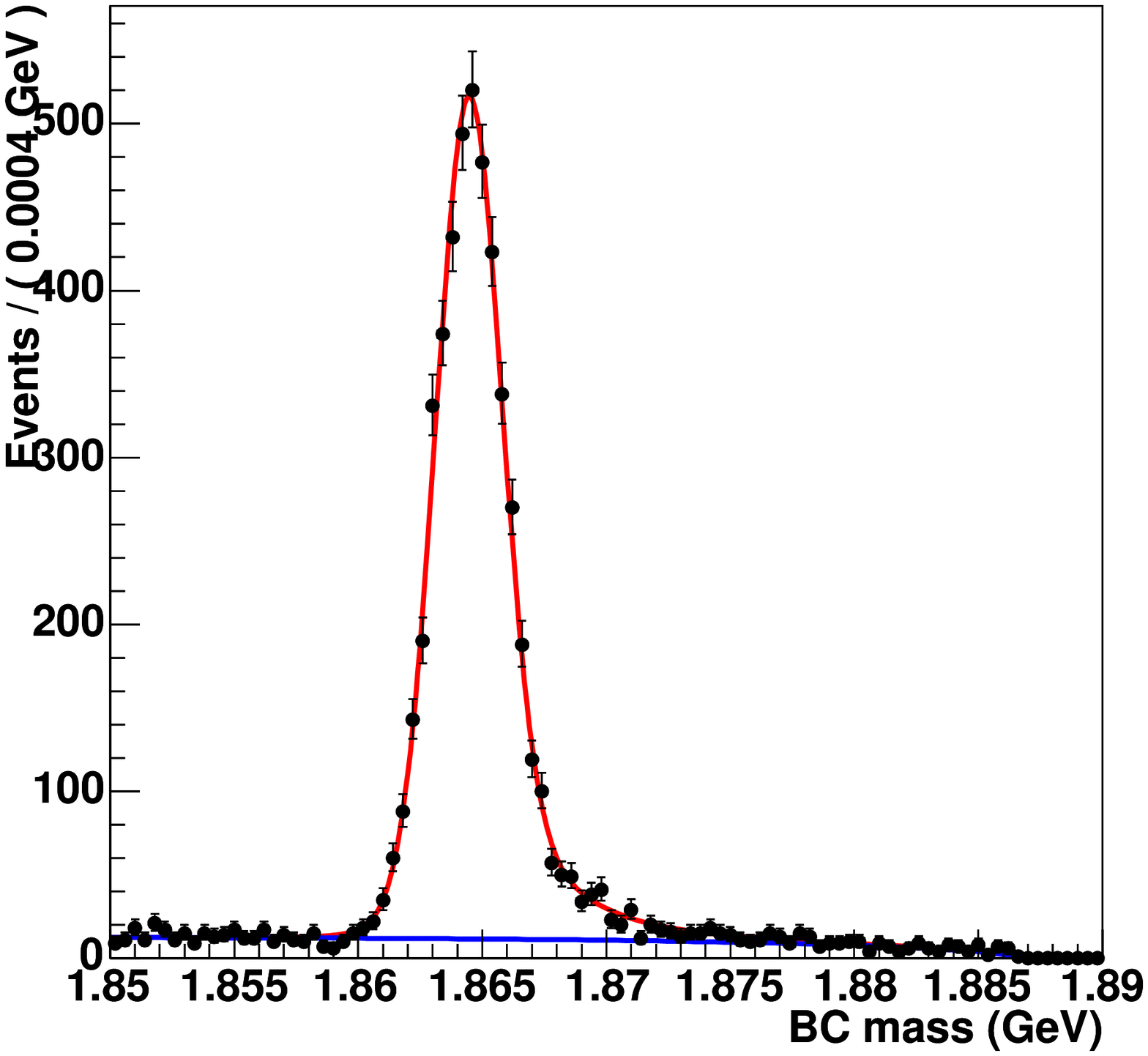}
\includegraphics[width=0.32\linewidth]{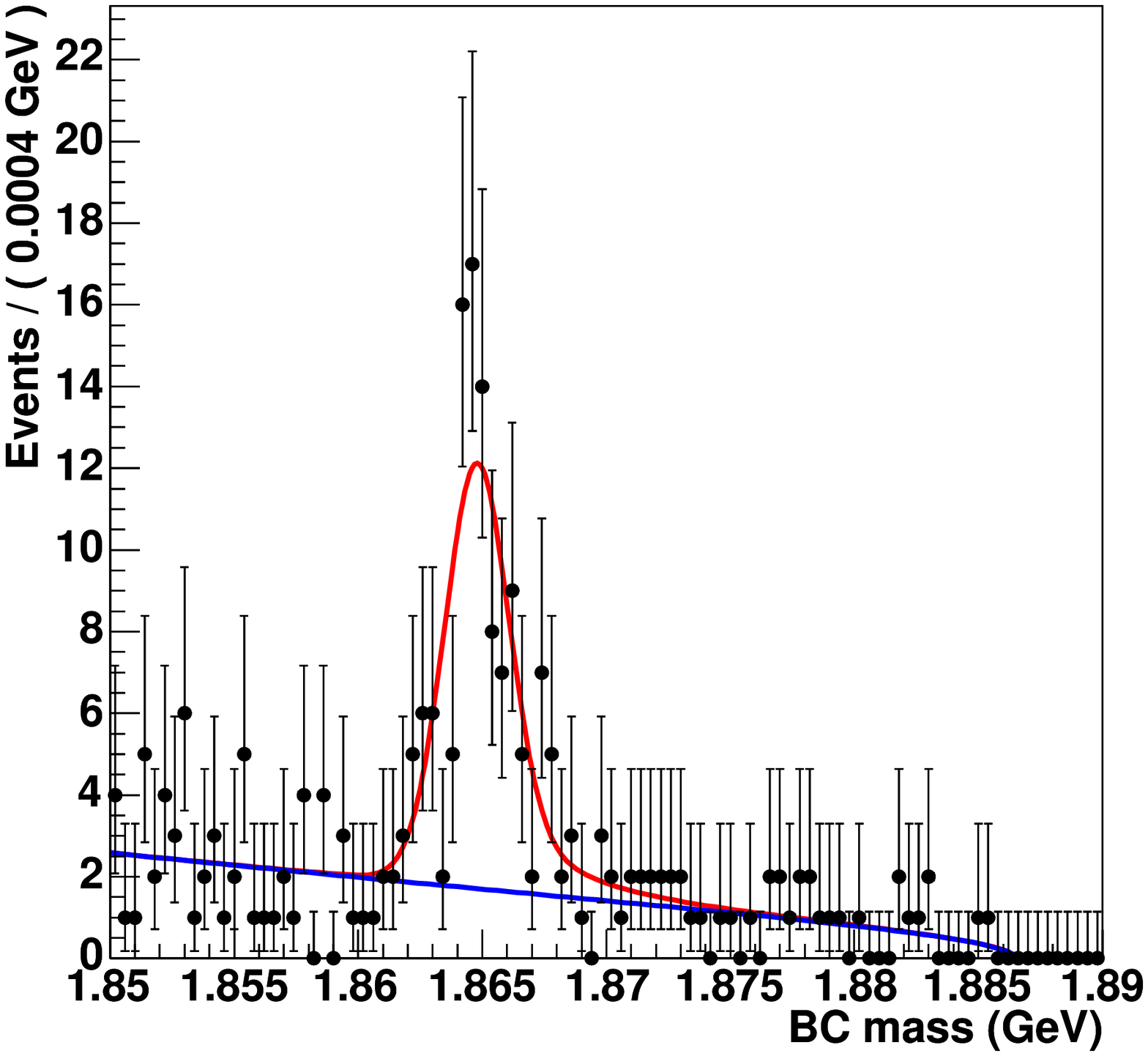}
\includegraphics[width=0.32\linewidth]{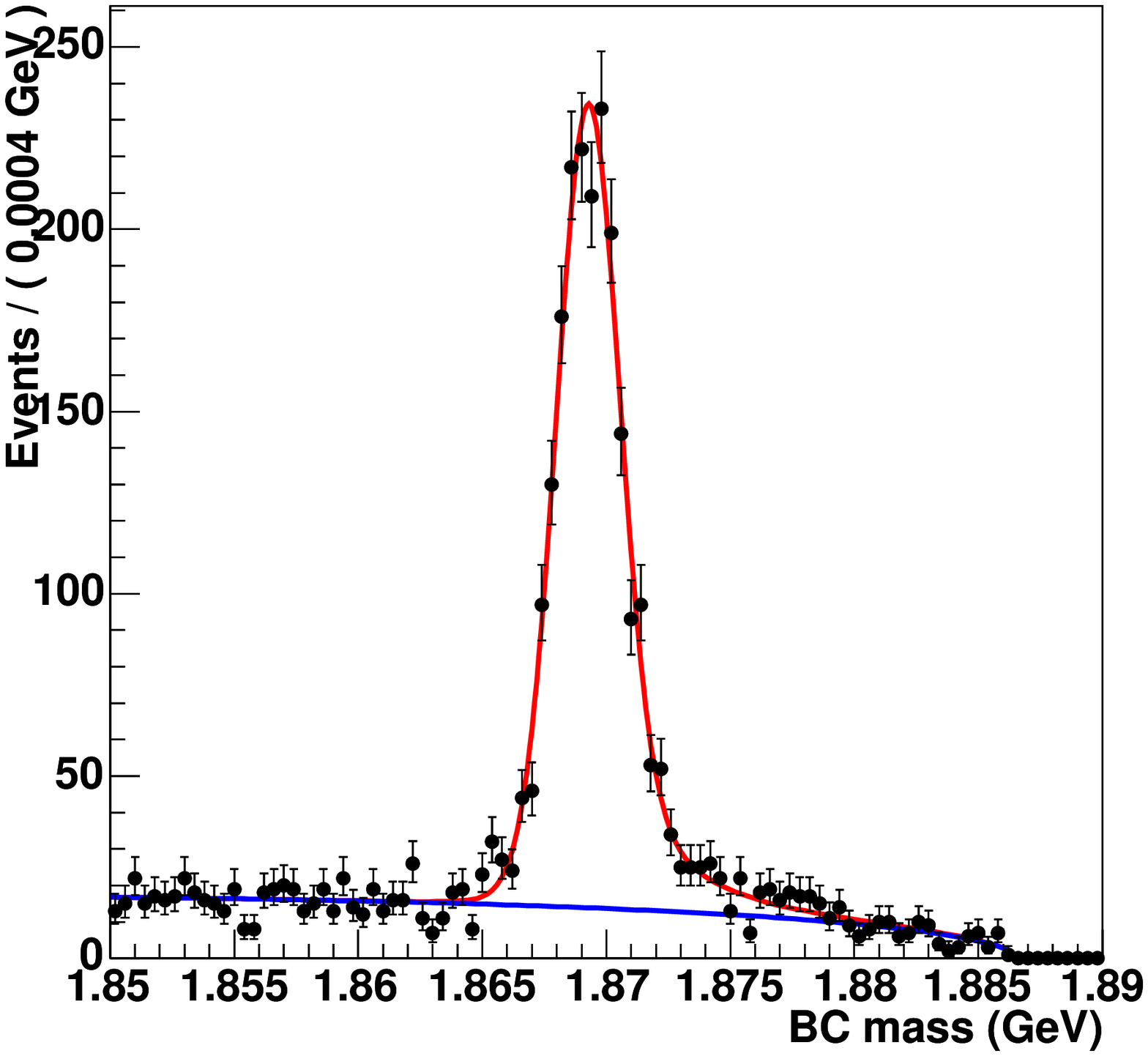}
\caption{From left to right the yields in the $D^0\to K^+K^-$,
$D^0\to K^0_SK^0_S$, and $D^+\to K^+K^0_S$ are shown. We observe
$4747\pm74$, $96\pm13$, and $1971\pm51$ events respectively in 
these modes. For the $D^0\to K^0_SK^0_S$ analysis we subtract
backgrounds, primarily, from $D^0\to K^0_S\pi^+\pi^-$ and 
find $70\pm15$ signal events.}
\label{fig:dtokk}
\end{center}
\end{figure*}

CLEO-c has studied Cabibbo suppressed two-body decays 
of $D^0$ and $D^+$ mesons to a pair of kaons. In particular,
the decays $D^0\to K^-K^+$,  $D^0\to K^0_SK^0_S$,
and $D^+\to K^+K^0_S$ have been analyzed. In addition to 
being Cabibbo suppressed,
the $D^0\to K^0_SK^0_S$ mode is strongly suppressed due to 
destructive interference in the SU(3) limit between the two 
dominating exchange amplitudes for this decay.
Figure~\ref{fig:dtokk} shows the observed yields in the 
three channels studied in this analysis. 

The preliminary branching fractions are summarized in Table~\ref{tab:dtokk}.
For $D^0\to K^+K^-$ and $D^+\to K^+K^0_S$ there is good agreement
with previous measurements. However, for $D^0\to K^0_SK^0_S$ our
new measurement is lower than previous measurements. 

\begin{table*}[bt]
\caption{Preliminary branching fractions obtained in the study of two-body
Cabibbo suppressed decays of $D$ mesons to pairs of kaons. }
\label{tab:dtokk}
\begin{center}
    \begin{tabular}{ccc} \hline\hline
    & Our Measurement ($10^{-3}$) & PDG 2007 ($10^{-3}$) \\ \hline
    ${\cal B}(D^{0} \to K^{-}K^{+})$         & $4.01 \pm 0.07 \pm 0.08 \pm 0.07$
 & $3.85 \pm 0.09 $  \\ 
    ${\cal B}(D^{0} \to K^{0}_{S}K^{0}_{S})$ & $0.149 \pm 0.034 \pm 0.015 \pm 0.
03 $ & $0.36 \pm 0.07 $  \\
    ${\cal B}(D^{+} \to K^{0}_{S}K^{+})$     & $3.35 \pm 0.10 \pm 0.10 \pm 0.12$
 & $ 2.95 \pm 0.19$  \\ \hline\hline
    \end{tabular}
\end{center}
\end{table*}

\section{Summary}

 I have presented results based on 281 pb$^{-1}$ of $e^+e^-$ annihilation
data recorded at the $\psi(3770)$ resonance for studies of $D^0$ and
$D^+$ decays. Among the results presented here were the final results
for the absolute $D^0\to K^-\pi^+$ and $D^+\to K^-\pi^+\pi^+$ 
branching fractions. CLEO-c has also analyzed 298 pb$^{-1}$ of $e^+e^-$
annihilation data recorded at the center-of-mass energy of 4170 MeV. 
Here we have studied the absolute hadronic branching fractions of
$D_s$ mesons. CLEO-c has recorded more than 800 pb$^{-1}$ 
of data at the  $\psi(3770)$
and are planing to double the data sample recorded at $E_{\rm cm}=4170$ MeV,
so there are still many interesting results to come from the CLEO-c 
data sample.

\section*{Acknowledgments}
This work was supported by the National Science Foundation grant
PHY-0202078 and by the Alfred P.~Sloan foundation.

\bigskip 

\end{document}